# Optimal photonic crystal cavities for coupling nanoemitters to photonic integrated circuits

*Jan Olthaus[1], Philip P. J. Schrinner[2], Doris E. Reiter[1*], Carsten Schuck[2*]*

[1]Institute of Solid State Theory, University of Münster, Wilhelm-Klemm-Str. 10, 48149 Münster, Germany
[2]Physics Institute, University of Münster, Wilhelm-Klemm-Str. 10, 48149 Münster, Germany
[*]E-mail: doris.reiter@uni-muenster.de, carsten.schuck@uni-muenster.de

Photonic integrated circuits that are manufactured with mature semiconductor technology hold great promise for realizing scalable quantum technology. Efficient interfaces between quantum emitters and nanophotonic devices are crucial building blocks for such implementations on silicon chips. These interfaces can be realized as nanobeam optical cavities with high quality factors and wavelength-scale mode volumes, thus providing enhanced coupling between nano-scale quantum emitters and nanophotonic circuits. Realizing such resonant structures is particularly challenging for the visible wavelength range, where many of the currently considered quantum emitters operate, and if compatibility with modern semiconductor nanofabrication processes is desired. Here we show that photonic crystal nanobeam cavities for the visible spectrum can be designed and fabricated directly on-substrate with high quality factors and small mode volumes. We compare designs based on deterministic and mode-matching methods and find the latter advantageous for on-substrate realizations. Our results pave the way for integrating quantum emitters with nanophotonic circuits for applications in quantum technology.

**Introduction**

Integrated quantum optics has a broad range of applications in quantum technologies [1, 2], including quantum information processing [3], and sensing [4]. One of the key building blocks for the realization of quantum technologies are non-classical light sources based on single photon emitters (SPEs). In the last decade, integrated quantum photonics has emerged as a tool to improve the performance of SPEs by altering their emission characteristics through integration with nanophotonic devices. The use of optical resonators is especially useful for improving the photon emission rate of SPEs via the Purcell effect [5-8]. This requires optical structures that support high-quality (Q)-factor resonator modes with small, i.e. wavelength-scale, mode volumes (Vm) tailored to the emission wavelength of a SPE. Photonic crystal (PhCs) cavities have shown to be the superior resonator choice for integrated optics as they provide high Q/Vm ratios [9, 10] and therewith enable efficient light-matter interfaces. PhC cavities can be implemented as two-dimensional (2D) slab geometries or one-dimensional (1D) nanobeams [11, 12], where the latter has a smaller device footprint while achieving similar Q-factors, thus favoring densely integrated photonic circuitry.

While several types of SPEs are considered for applications in quantum optics [13], e.g. quantum dots [14, 15] or carbon nanotubes [16, 17], nitrogen vacancy (NV)-centers in diamond are particularly promising candidates because they are photostable and feature long spin-coherence times [18]. In integrated optics a great interest in coupling NV-centers to photonic crystal cavities ensued [8, 19-23] in order to make their attractive photophysical properties available in nanophotonic networks. Corresponding nanophotonic devices need to be fabricated from material systems, which are transparent in the visible wavelength range, in particular at 532 nm and 637 nm where NV-centers can be excited and emit fluorescence, respectively. Here we use silicon nitride (SiN) because of its excellent compatibility with nanofabrication processes developed for high-quality photonic integrated circuits. While the devices considered in our work are optimized for NV-centers in nanodiamonds, similar coupling strategies apply for other quantum emitters, e.g. colloidal quantum dots or single-molecules, in nanoscale host material volumes that can be positioned in close proximity of a SiN nanobeam cavity.

Most of the SiN resonator designs reported in the literature consider free-standing PhC nanobeams [23-25], i.e. the substrate below the nanobeam is removed. Free-standing nanobeam cavities benefit from high refractive index contrast between the waveguide and a uniform environment (air) for achieving high Q-factors. However, such devices have limited compatibility with semiconductor industry processes, which will be desirable for future large-scale fabrication of integrated quantum technology. In an effort to avoid free-standing geometries but nevertheless provide uniform environments around a PhC-cavity such devices have been encapsulated with poly(methyl methacrylate) (PMMA) [7, 26] resulting in Q-factors of up to $10^5$ in simulations and several thousands in fabricated devices for 764 nm wavelength. Here we show how SiN-PhC nanobeam cavities with high Q/Vm ratios can be realized directly on a silicon dioxide ($SiO_2$) substrate to provide efficient interfaces to single quantum emitters in the visible wavelength range. Our designs do not require additional processing steps that have limited compatibility with established thin-film technology for large-scale photonic circuit integration. A further advantage of our on-substrate designs is their improved thermalization with the substrate, which strongly benefits low-temperature implementations of waveguide-integrated quantum emitters that typically suffer from low specific heat capacity of suspended structures [27].

We find optimal performance for on-substrate PhC-cavities by performing comparative studies of deterministic and mode-matching designs both in numerical simulation and experimental realization. The deterministic design approaches are based on band structure simulations [10, 28], while the mode-matching approach consists of brute-force parameter optimization [23, 29].

We explicitly take fabrication constraints into account and consider that differences in hole size and distance (mode-matching design) as well as changes in waveguide width (deterministic designs) in the taper section have the strongest influence on device performance. We show that for a fixed number of holes, the mode-matching design requires a smaller taper section than the deterministic designs to obtain comparable Q-factors, which is advantageous in device fabrication. Based on the simulation results we fabricate integrated PhC devices optimized for 637 nm wavelength, i.e. the zero phonon line of the NV⁻ center in diamond, and measure resonances with Q-factors exceeding $10^3$.

**Cavity design**

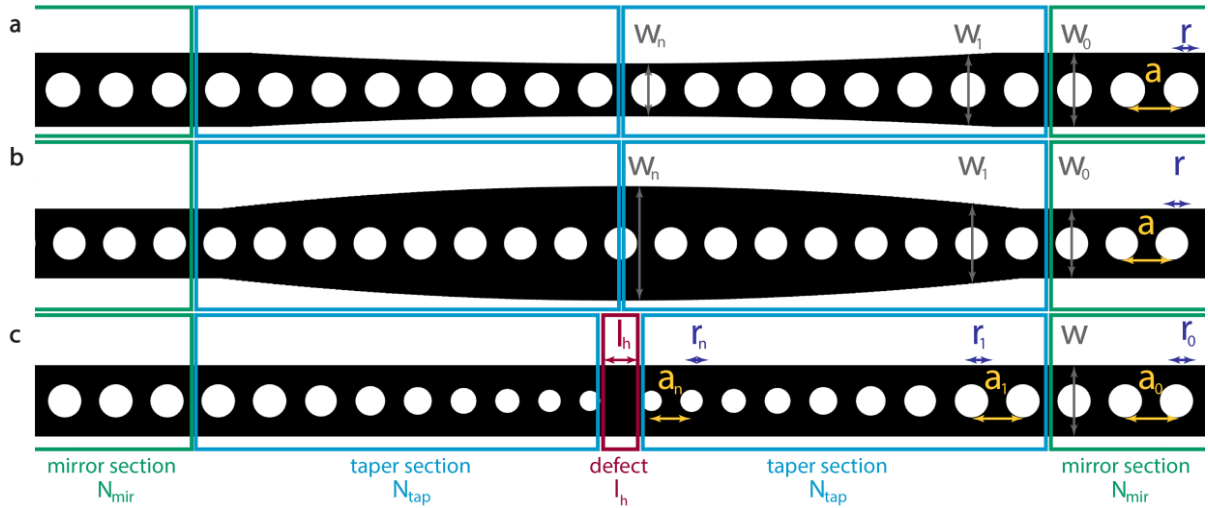

**Figure 1: Sketch of the PhC nanobeam cavity designs.** A PhC cavity consists of a mirror section with $N_{\text{mir}}$ holes and a taper section with $N_{tap}$ holes on either side, resulting in a confined mode at the center. **a/b** Deterministic designs, where the cavity mode is created by changing the width of the waveguide, while the hole size and distance stays constant. **c** Mode-matching design, where the radii of the holes and the hole distances are linearly variied and a defect of length $l_h$ is inserted.

We consider design approaches for PhC-cavities where mirror sections consisting of $N_{mir}$ periodic holes, as shown in Figure 1, create a photonic band gap around the resonant frequency $\omega_{res}$. These mirror sections have to be optimized in terms of the mirror strength $\gamma$ to achieve high quality factors.

In the deterministic design approaches, originally developed in ref. [28],[10], the confined mode in the band gap is created by quadratic tapering of the waveguide width, $w$. In Figure 1 a the central waveguide width is decreased resulting in the dielectric mode moving into the bandgap ($\varepsilon$-mode design). In Figure 1 b instead, the central waveguide width is increased such that the air-mode moves into the bandgap (air-mode design). These approaches minimize the out-of-plane scattering by creating an attenuation profile of Gaussian-shape for the defect mode along the taper section.

Figure 1c shows the mode-matching design approach [29]. In this case the hole radii $r$ and hole distances $a$ are tapered down linearly from the mirror sections to the defect center, created by the defect length $l_h$. This maximizes the overlap integral of the modes within two neighboring segments by gradual variation of the band structure, thus resulting in minimal loss. Note that in the deterministic design no defect length $l_h$ is required; instead the mode is concentrated around the interface between the two taper sections.

**Periodic 1D PhC nanobeam/band structure analysis**

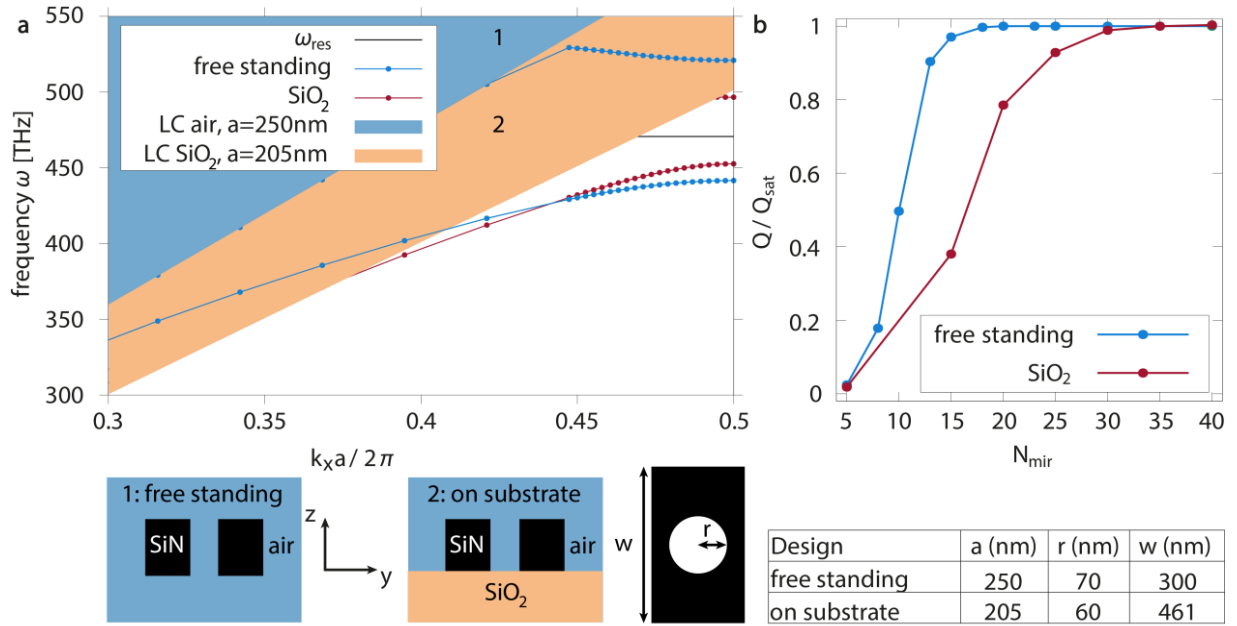

**Figure 2: TE-Band structure analysis. a** $\varepsilon$- and air-modes (dotted lines) for periodic free standing (blue) and on-substrate (red) PhC nanobeam waveguides in the 1.BZ. The shaded regions represent the light cones for air and SiO$_2$ substrate and are only relevant for the respective designs. The targeted resonance frequency is shown as black line. The bottom of the figure shows the schematic unit cells along

the parameters for optimized mirror strength. **b** Variation of Q-factor with $N_{mir}$ for $N_{tap}= 10$ for a mode-matching design. Analyzed for free-standing (blue) and on-substrate (red) devices. The Q-factors are normalized by the respective saturated values $Q_{sat}$.

In a first optimization step we employ the frequency-eigensolver MPB [30] to calculate the band structure for mirror sections that are defined by the unit cell shown in Figure 2a, with hole distance $a$, hole radius $r$ and waveguide width $w$. We fix the thickness of the SiN waveguide at $t = 200$ nm to match fabrication constraints. We then maximize the mirror strength $\gamma$ by variing $a$, $r$ and $w$, taking into account that the target frequency $\omega_{res}$ (Figure 2, black line) should lie in the center of the bandgap. The resulting parameters for the on-substrate design are $a = 205$ nm, $r = 60$ nm and $w = 461$ nm. (The unit cell used for the mode-matching approach has slightly deviating mirror parameters: $a = 205$ nm, $r = 56$ nm and $w = 492$ nm, resulting in a slightly lower $\gamma$). We compare this on-substrate PhC with a similar optimized free-standing PhC where parameter optimization yields $a = 250$ nm, $r = 70$ nm and $w = 300$ nm, similar to previous results [23]. The respective band structure calculations for the free standing (blue) and on-substrate designs (red) are shown in Figure 2a.

Optimal designs for free-standing and on-substrate nanobeams differ most significantly in their hole size and waveguide width. A smaller hole size for optimal on-substrate geometries as compared to free-standing designs is reasonable because the effective refractive index inside the unit cell is higher. Similarly, larger waveguide width for optimal on-substrate geometries as compared to free-standing designs are expected because a larger amount of material (SiN) is needed to confine the mode inside the waveguide rather than in the underlying substrate ($SiO_2$). We note that the resulting bandgap is located rather close to the respective light cone. A further increase of the waveguide width would separate the band frequencies further from the light cone, but also result in decreased mirror strength $\gamma$ due to a smaller size of the bandgap.

We find that the bandgap of the on-substrate design is significantly smaller than that of the free-standing design. This indicates that out-of-plane scattering into the substrate leads to reduced quality-factor and hints at the expected adverse effect of a substrate on the Q-factor.

After optimizing the unit cell of the mirror segments, we determine the number of mirror holes, for which the PhC losses are limited by out-of-plane scattering. An increasing number of mirror holes, $N_{mir}$, results in reduced transmission through the waveguide, thus impacting the signal-to-noise ratio in a measurement. Figure 2b shows the relation between $N_{mir}$ and the Q-factor for mode-matching designs with $N_{tap}= 10$, both for free-standing (cf. Ref. [23]) and on-substrate PhC nanobeams. While for the free-standing case a saturation of the Q–factor is reached already for $N_{mir} > 17$, in the on-substrate case Q is only saturated for $N_{mir} > 30$. This difference is expected as a consequence of the lower mirror strength $\gamma$ in the on-substrate case.

**1D PhC nanobeam cavity**

In a second step we optimize the taper section of the PhC cavities and analyze their Q-factors. Here we use the optimal mirror sections calculated in the previous section as a starting point for performing 3D-finite difference time domain (FDTD) simulations with the MEEP software package [30].

For our simulations we fix the total number of holes to $N_{tot} = 35$ for all three designs and vary the number of taper hole $N_{tap}$. This choice allows for comparing different designs and ensures sufficient transmission in experimental measurements (see below). We further perform simulations for a fixed number of mirror holes, $N_{mir} = 35$, to analyze the influence of transmission losses.

In the deterministic designs quadratic tapering of the waveguide width creates a defect mode. We first tune the defect mode frequency to approximately $\omega_{res}$ for $N_{tap} = 35$ through variation of the central waveguide width $w_n$ (see Figure 1). An advantage of the deterministic designs is that results of the band structure calculations can be exploited for estimating the central waveguide width $w_n$ by comparing the variations of the $\varepsilon$-mode and air-mode frequencies at the edge of the first Brillouin zone when varying the waveguide width [10]. However, these values need to be verified by 3D-FDTD simulations because the exact deviation between the mode frequencies of the central segment and the resulting defect mode frequencies depends on the number of taper holes, $N_{tap}$. We find the central waveguide width $w_n = 625$ nm for the air-mode design and $w_n = 338$ nm for the $\varepsilon$-mode design. Based on these values we then calculate the quality factor of the defect mode under variation of $N_{tap}$.

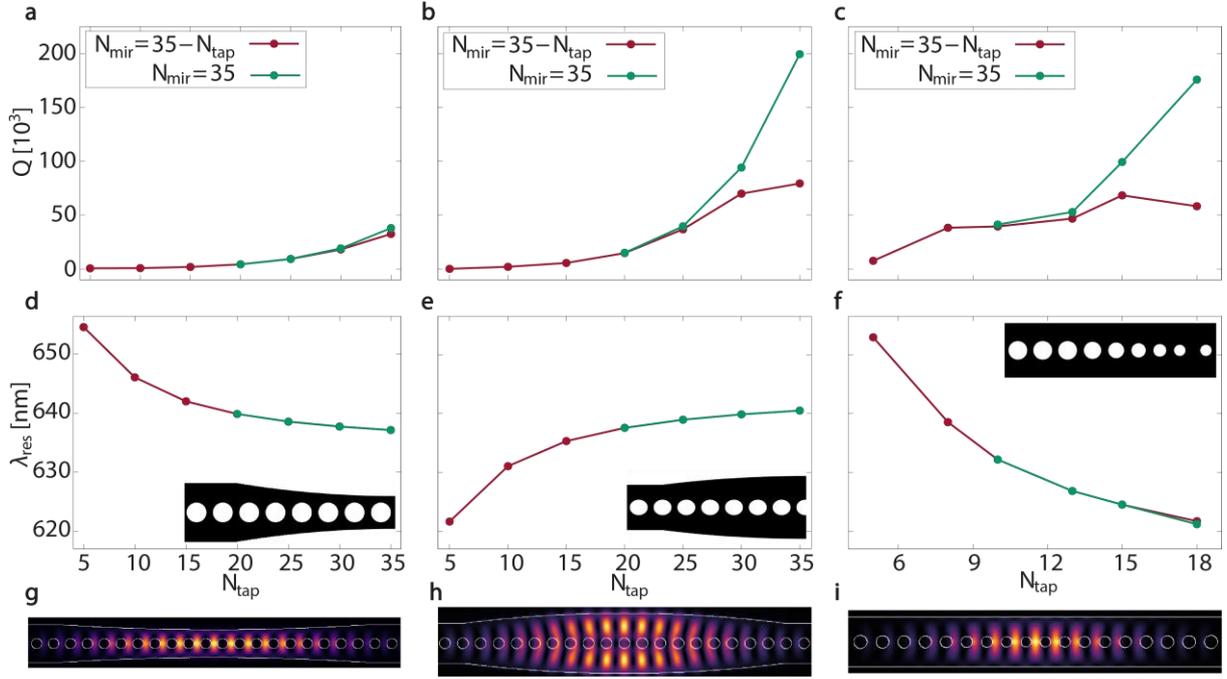

**Figure 3: Simulation of three PhC cavity designs. a/b/c** Q-factors and **d/e/f** resonant wavelengths $\lambda_{res}$ as function of $N_{tap}$ for unsaturated ($N_{tot} = 35$, red curves) and saturated ($N_{mir} = 35$, green curves) PhC cavities for the three design approaches. **g/h/i** mode profiles in the x-y plane.

The resulting Q-factors and the resonance wavelengths $\lambda_{res}$ are show in Figure 3 a and d for the $\varepsilon$-mode design and in Figure 3 b and e for the air-mode design, respectively. Exemplary field profiles for the corresponding resonator modes are depicted in Figure 3 g and h, respectively. For both designs, we see a continuous increase of the Q-factor with $N_{tap}$. This results in Q-factors of $3.3 \cdot 10^4$ for the $\varepsilon$-mode design and $7.9 \cdot 10^4$ for the air-mode design at the maximal number of taper holes $N_{tap} = 35$. For the $\varepsilon$-mode the resonant wavelength $\lambda_{res}$ decreases with $N_{tap}$ while it increases in the case of the air-mode PhC cavity (Figure 3 d,e). This behavior is expected as the defect constitutes a perturbation of the optimized mirror segment (see Figure 2 a), which supports an $\varepsilon$-mode with wavelengths $> \lambda_{res}$ and an air-mode with wavelengths $< \lambda_{res}$. Both designs approach the target wavelength of $\lambda = 637$ nm at $N_{mir} = 35$.

By increasing the overall number of holes we find that for $N_{tap} > 20$ both cavities are not completely scattering-limited anymore. In the case where the number of mirror holes, $N_{mir}$, is large enough to saturate the Q-factor, i.e. we switch off the transmission losses, we find Q-factors of $3.8 \cdot 10^4$ for the $\varepsilon$-mode design and $2.0 \cdot 10^5$ for the air-more design at the highest simulated taper hole number of $N_{tap} = 35$, as shown in Figure 3 a and b, respectively.

We further extract the mode volumes from our calculations. With increasing $N_{tap}$ the mode volume for the $\varepsilon$-mode design rises linearly from $V_m = 1.29 \left(\frac{\lambda_{res}}{n_{SiN}}\right)^3$ at $N_{tap} = 5$ to $V_m = 2.30 \left(\frac{\lambda_{res}}{n_{SiN}}\right)^3$ at $N_{tap} = 35$, while for the air-mode design we find an increase from $V_m = 2.45 \left(\frac{\lambda_{res}}{n_{SiN}}\right)^3$ at $N_{tap} = 5$ to $V_m = 5.98 \left(\frac{\lambda_{res}}{n_{SiN}}\right)^3$ at $N_{tap} = 35$. The significantly higher mode volumes of the air-mode design as compared to the $\varepsilon$-mode design are also visible in the mode profiles shown in Figure 3 g and h.

When considering the coupling of a NV-center in a nanodiamond to the waveguide, it is noteworthy, that the air-mode design might be favorable, because its field maximum is located inside the air-hole regions, where nanodiamonds could be placed. In contrast, in the $\varepsilon$-mode design, the field maximum is concentrated inside the dielectric medium such that emitters need to be placed inside the evanescent field surrounding the structure.

We now compare these deterministic designs to the mode-matching design where we are tapering hole distances and diameters as well as tuning the defect length, $l_h$. The resulting Q-factors are shown in Figure 3 c, the resonance wavelengths in Figure 3 f and an exemplary mode profile in Figure 3 i. In this design, the defect length is used to tune the cavity frequency. Accordingly, the Q-factor depends strongly on the defect length $l_h$, which needs to be adjusted for each $N_{tap}$, with optimal values between $l_h = 80 - 90$nm (We discuss the optimization of $l_h$ in more detail in the supplementary information (SI), see Figure S1).

The optimization of the Q-factors in the mode-matching design relies on varying the size of the central segments $a_n$ and the radii of the central holes $r_n$, which are both changed linearly with respect to $a_0$ and $r_0$ of the mirror section (see Figure 1). For the mode-matching design, the taper optimization is computationally very expensive, because prior parameter estimation from the design approach are not possible (except for an approximation of the defect length $l_h$ to match the target frequency). The parameters $a_n$ and $r_n$ have to be iteratively optimized in terms of both the Q-factor and wavelength $\lambda_{res}$ by executing 3D-FDTD simulations of the entire cavity structure. Using this brute force approach we find optimal design parameters for the taper section as $a_n = 175$ nm and $r_n = 46$ nm. A resulting mode profile is shown in Figure 3i.

Based on this parameter set we calculate the Q-factor as a function of the number of taper holes $N_{tap}$ as shown in Figure 3c. Note that the x-axis in this case only shows maximal values of up to $N_{tap} = 18$ (as compared to $N_{tap} = 35$ for the deterministic designs) and that the Q-factors depicted here are always the result of cavities with an optimal defect length $l_h$ (see SI). We find that the quality factor reaches a maximal value of $Q = 6.8 \cdot 10^4$ at $N_{tap} = 15$, where the mode volume $V_m = 1.41 \left(\frac{\lambda_{res}}{n_{SiN}}\right)^3$. If $N_{tap}$ is increased further, the Q-factor decreases because transmission losses become dominant. This is seen from comparison with saturated Q-factors for large numbers of $N_{mir}$ (green curve) reaching $Q = 1.75 \cdot 10^5$ with a corresponding $V_m = 1.48 \left(\frac{\lambda_{res}}{n_{SiN}}\right)^3$ at $N_{tap} = 18$.

From our simulations we find that the Q-factors for the deterministic air-mode design and the mode-matching design reach comparable values, however, the latter requiring a much smaller number of taper holes ($N_{tap} = 15$ for the mode-matching design vs. $N_{tap} = 35$ for the air-mode design). The mode-matching design is further favorable for realizing small mode volumes, i.e. less than one quarter of the mode volume for the air-mode design. The mode volume the $\varepsilon$-mode design on the other hand is comparable to that of the mode-matching design, but features much lower Q-factors. Overall, we conclude that the mode-matching design shows favourable performance for realizing efficient interfaces to quantum emitters, such as the NV-centers considered here, because it achieves maximal Q/Vm ratios for a low number of taper holes.

We further find that Q-factors $> 10^6$ are achievable with free-standing SiN PhC designs for even smaller taper sections ($N_{tap} > 7$), while maintaining a sub-wavelength Vm (not shown). In such free-standing designs loss channels into the substrate are strongly suppressed and significantly higher Q/Vm ratios are achievable (in simulation) [24]. However in experiments targeting the visible wavelength range it turns out to be difficult to take advantage of the theoretical performance of free-standing designs, which are not compatible with current semiconductor industry processes for realizing large scale photonic integrated circuit implementations.

**Experimental Results**

In order to experimentally test the simulated PhC-cavity designs and the conclusions drawn from the simulated performance characteristics we fabricate PhC nanobeam cavities with deterministic dielectric-mode and air-mode designs as well as mode-matching designs. The cavities are integrated with nanophotonic waveguides that allow for performing transmission measurements from which we extract the quality factors and resonance wavelengths.

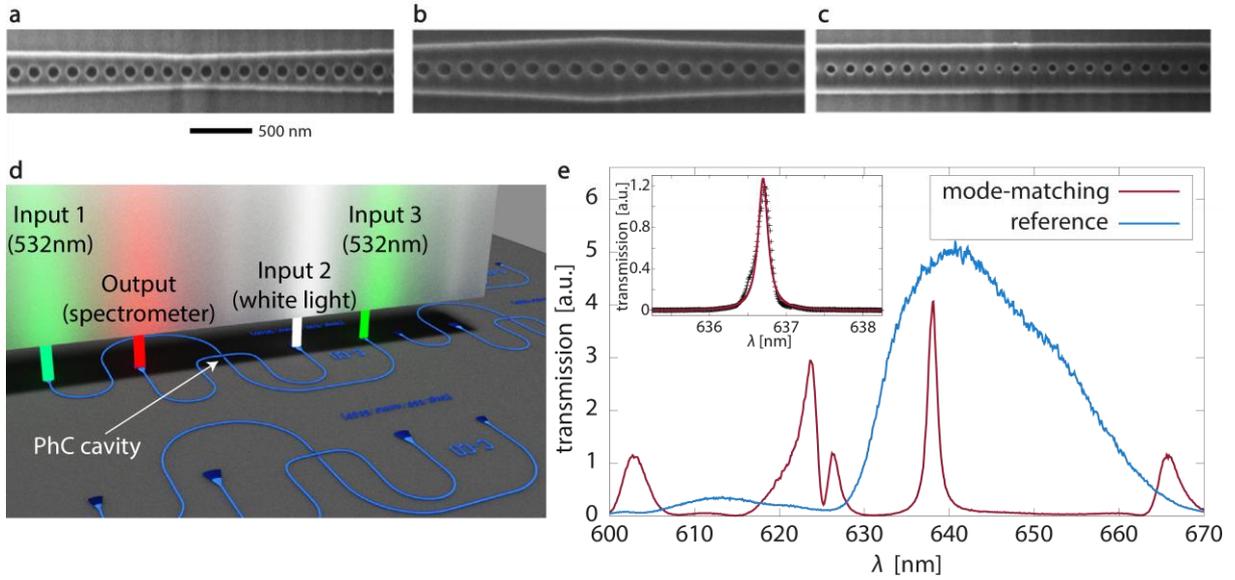

**Figure 4: Fabrication, setup and measurement of the PhC cavities devices.** SEM images of the fabricated PhC cavities with **a** ε-mode, **b** air-mode and **c** mode-matching design (scale-bar: 500 nm). **d** Schematic of the transmission measurement setup. **e** Transmission spectrum for a mode-matching PhC cavity device with 6 taper holes and 15 mirror holes on substrate (red) and a reference device (blue) with grating couplers only (scaled for better comparison). The measured bandgap spans from 603 – 665 nm with a resonance at the desired target wavelength of 637 nm. We attribute the artifact at 617-627 nm to undesired TM modes. The peak at 637 nm is the resonance of the cavity. Inset: Resonance of a PhC cavity with a total of 34 holes. Lorentzian fitting yields a Q-factor of $4187 \pm 11$.

The PhC nanobeam cavity devices are fabricated from 200 nm SiN thin films on insulator using electron beam lithography (EBPG) and reactive ion etching (see SI for details). Each nanobeam cavity is fabricated directly on the SiO2 substrate and connected via nanophotonic waveguides to optical grating couplers that provide interfaces to optical fibers. We optimize our nanofabrication recipes to accurately produce the design values of each PhC nanobeam cavity as variations in waveguide width and hole radius can lead to significant shifts of the resonance wavelength and reduced Q-factors [31]. We assess the fabrication tolerances of the devices in scanning electron microscopy (SEM), revealing overall low side-wall roughness and high circularity of the holes, as shown in Figure 4 a)-c). We are able to match the simulated mode-matching and air-mode design values for waveguide width and hole radius in fabricated devices to within approximately 10 nm and 3 nm, respectively. For the dielectric-mode design we find similar tolerances for the waveguide width, but the hole radii in the narrow center region are increased by approximately 6 nm. This labels the dielectric-design somewhat more challenging in fabrication as compared to the air-mode and mode-matching designs.

We find that fabricated devices with large numbers of holes suffer from low optical transmission and hence only considered device designs with $N_{tot} \leq 35$ holes. This choice allows for simultaneously obtaining high signal-to-noise transmission measurement signatures and high quality-factors, in accordance with above FDTD simulations. We restrict our experimental studies to device designs with parameter sets for which we expect the highest quality factors from Figure 3, because each fabricated design requires fine-tuning of the defect length, $l_h$, hole distance, $a$, hole radius, $r$, and waveguide width, $w$, in order to achieve resonances at the desired target wavelength (i.e. 637 nm). Consequently we consider mode-matching geometries with $N_{tap} \leq 15$ taper holes and deterministic design with $N_{tap} = 20$-35 taper holes.

We assess the Q-factors and resonance wavelengths of fabricated PhC cavity devices in transmission measurements shown schematically in Figure 4d. The PhC nanobeam cavity is located at the center of the nanophotonic circuit and connects to 500 nm wide, 200 nm thick waveguides that allow for supplying white light from a polarized supercontinuum laser source (450-900 nm) via optical grating couplers aligned to input 2 of a single-mode fiber array (see Figure 4d). Light that is transmitted through the PhC nanobeam cavity is guided via a waveguide of similar dimensions to a grating coupler aligned with the output fiber in the array that connects to a spectrometer. The grating couplers are optimized for transverse electric (TE) optical modes at the desired target wavelength and support a -3dB bandwidth of ~20 nm as shown by the blue curve in Figure 4e. This measurement configuration enables measurements of the entire bandgap when considering the 80 nm band over which the transmission from a grating coupler exceeds the noise floor of the spectrometer.

In anticipation of integrating fluorescent nanoemitters with the PhC cavities the devices shown in Figure 4d feature additional optical access optimized for guiding 532 nm light from inputs 1 and 3 to the center of the nanobeam under perpendicular incidence. In the future these input ports will allow for optical excitation of NV-centers in diamonds positioned within (the vicinity of) the mode volume of the cavity but do not influence the cavity performance in the designs considered here. Modern foundry services may also enable innovative emitter-waveguide coupling strategies by exploiting advanced multi-layer thin-film processes [32].

Transmission measurements from fiber-input 2 to the output-fiber (Figure 4d) show characteristic spectra similar to the one presented in Figure 4e for a PhC cavity with mode-matching design and $N_{tot} = 21$ holes. For optimized designs we observe the edges of the photonic bandgap at 603 nm and 665 nm as well as a resonance peak at the desired target wavelength of 637 nm. Within the bandgap transmission is suppressed by -30 dB with respect to the transmission on resonance. We further observe artifacts in the spectrum (here at 617-627 nm), which we attribute to transverse magnetic (TM) modes that are guided in our 200 nm thick waveguides and lie outside the design bandwidth of our grating couplers.

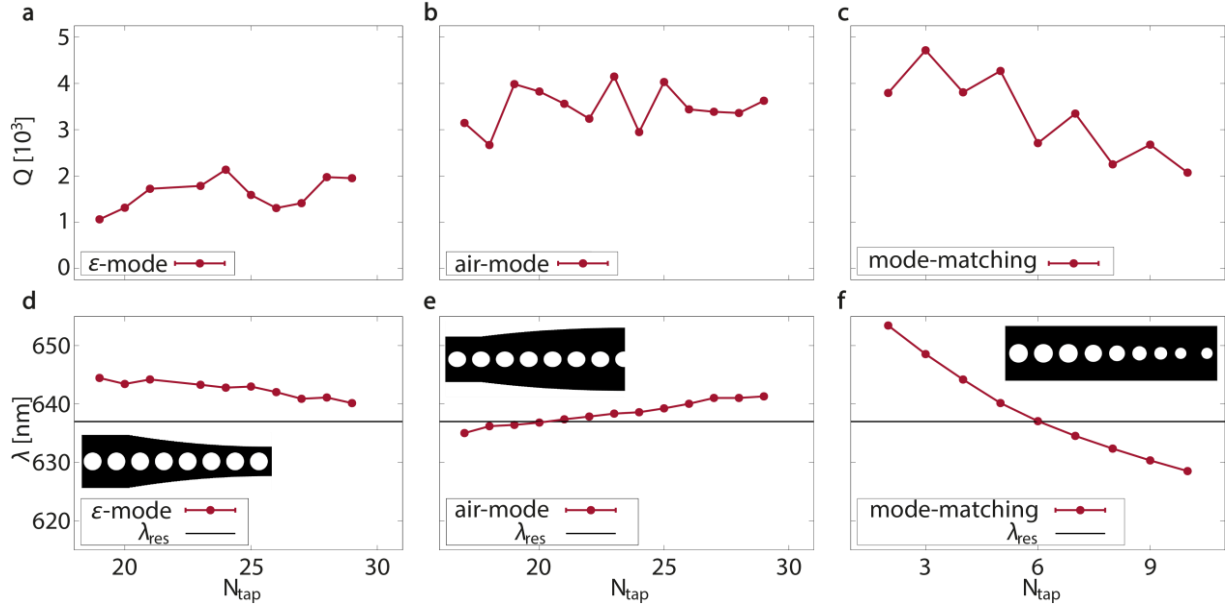

**Figure 5: Measurements of the three PhC cavity geometries. a/b/c** Q-factors and **d/e/f** resonance wavelengths $\lambda$ for different numbers of taper holes $N_{tap}$ and a total number of $N_{tot} = 35$ holes for the $\varepsilon$-mode, air-mode and mode-matching designs, respectively.

For deterministic PhC cavity designs with $N_{tot} = 35$ holes we observe resonances with quality-factors of several thousands for all considered numbers of taper holes as shown in Figure 5a and b. The Q-factors extracted from measured devices generally fall below those found in FDTD simulations but confirm the conclusion that air-mode designs yield higher Q-factors as compared to dielectric-mode designs, as also seen in Figure 3a and b. The resonance wavelength for both dielectric- and air-mode designs, shown in Figure 5d and e, respectively, approaches the desired target wavelength in qualitatively similar fashion as expected from FDTD simulations (see Figure 3d and e).

For mode-matching designs we observe that the highest quality factors (up to Q = 4500) are realized for taper sections with 3-7 holes, as visible in Figure 5c for devices with $N_{tot} = 35$. We measure resonance wavelengths in a 20 nm band around the target wavelength of 637 nm for fixed defect lengths, as shown in Figure 5f. Devices with $N_{tap} > 10$ are not shown in Figure 5f because the bandwidth of our grating couplers limits transmission of cavity resonances in this wavelength range, as evident from Figure 4e. This circumstance is owed to the relatively strong shift of the resonance frequency with the number of taper holes for fixed defect lengths as compared to deterministic designs, which was also observed in the simulation data of Figure 3. In designs with 6 taper holes and 28 mirror holes we extract Q-factors of Q = 4187 ± 11 from a Lorentzian fit to the data at the desired target wavelength, as shown in the inset of Fig 4e. We note that the simulated Q-factor is larger than the measured Q-factor, which has additional loss contributions, e.g scattering due to fabrication imperfection and absorption in the substrate.

We conclude that material and fabrication imperfections (e.g. the resolution of our EBPG system) prevent the experimental realization of mode-matching designs with higher Q-factors for larger numbers of taper holes, resulting in maximal measured Q-values for $N_{tap} < 7$. Nevertheless, the mode-matching designs achieve the highest overall Q-factors for the lowest number of taper holes among all design types, which is consistent with our expectation from 3D-FDTD simulation results (see Figure 3). Fabrication imperfections are most severe for the dielectric-mode design, which also yielded the lowest Q-factors after parameter optimization. We find reasonable qualitative agreement when comparing dielectric-mode, air-mode and mode-matching designs in 3D-FDTD simulations and measurements of devices fabricated in electron beam lithography and reactive ion etching. Our results imply that PhC cavities with mode-matching designs are advantageous for realizing efficient interfaces between optical waveguides and nanoscale emitters, because they achieve high Q-factors already for a small number of taper holes, which are the most challenging constituent of PhC cavities in terms of nanofabrication tolerances.

**Conclusions and Outlook**

In summary, we have optimized SiN nanobeam PhC cavities on $SiO_2$ substrates in the visible wavelength range for the coupling of nanoscale emitters to photonic integrated circuits. As an exemplary use-case we consider NV-centers in diamond as quantum emitters and optimize PhC cavity designs for a target wavelength of 637 nm, corresponding to the zero phonon line of $NV^-$-defects. As feature sizes in resonant nanophotonic structures scale with (fractions of the) wavelength our use-case realizes particularly challenging nanofabrication requirements as compared to many other solid-state quantum sources emitting at longer wavelengths. We take these nanofabrication requirements into account throughout the simulation of suitable designs and show that it is possible to realize high-Q PhC cavities in SiN nanobeams despite the small refractive index contrast with the $SiO_2$ substrate. A comparison of three on-substrate cavity designs ($\varepsilon$-mode, air-mode and mode-matching) in 3D-FDTD simulations yields the highest

Q/Vm ratios for mode-matching designs that show Q-factors > $10^5$ while maintaining wavelength-scale mode volumes.

We test the feasibility of realizing efficient light-matter interfaces at visible wavelengths (here 637 nm) as on-substrate nanobeam PhC cavities in CMOS-compatible processes by fabricating such devices in state-of-the-art lithography on SiN thin-films. Through careful parameter tuning we are able to realize PhC cavity devices with resonance wavelengths of 637 nm for all three geometries ($\varepsilon$-mode, air-mode and mode-matching). We find that mode-matching designs with smaller taper sections, consisting of only 3-7 holes, show superior performance, i.e. Q-factors, over deterministic designs with significantly larger taper sections of up to 30 holes. These findings are in qualitative agreement with 3D-FDTD simulations while showing lower overall Q-factors, which we attribute to fabrication and material imperfections. Such imperfections are easier mitigated for mode-matching geometries, which feature a larger design parameter space as compared to deterministic geometries.

We envision further optimization of PhC cavity interfaces through modified geometries that compensate correlations between waveguide width and hole diameters, by considering elliptical hole shapes and by resorting to ridge or strip-loaded waveguide geometries [21]. Future experimental realizations should also suppress the TM modes observed in our transmission spectra, as these provide additional loss channels. This could for example be achieved by reducing the waveguide height, which would however also result in a smaller effective refractive index contrast between nanobeam and substrate and thus yield smaller band gaps. The integration of nanoscale emitters with nanobeam PhC cavities will further require methods for compensating the effects of scattering centers in the vicinity of the cavity mode volume.

We conclude that even moderate Purcell-enhancement, as expected for integrating nanoemitters with the PhC cavity geometries considered here, will notably enhance the coupling efficiency to waveguides. Nanophotonic circuits further offer great versatility in configuring the optical inputs and outputs to light-matter interfaces like the PhC nanobeam cavities presented here (see Figure 4d). Such integrated solutions are attractive for simultaneously interfacing photonic networks with large numbers of emitters in a scalable fashion by exploiting small device footprints and highly reproducible fabrication routines. Our results thus pave the way for supplying single-photons from a large number of nanoscale emitters into quantum photonic circuits as desired for integrated quantum technologies.

**Acknowledgements**

C.S. acknowledges support from the Ministry for Culture and Science of North Rhine-Westphalia.

# Supporting Information

**Optimal photonic crystal cavities for coupling nanoemitters to photonic integrated circuits**

*Jan Olthaus[1], Philip P. J. Schrinner[2], Doris E. Reiter[1*], Carsten Schuck[2*]*

[1]Institute of Solid State Theory, University of Münster, Wilhelm-Klemm-Str. 10, 48149 Münster, Germany
[2]Physics Institute, University of Münster, Wilhelm-Klemm-Str. 10, 48149 Münster, Germany
[*]E-mail: doris.reiter@uni-muenster.de, carsten.schuck@uni-muenster.de

**Optimization of mode-matching designs**

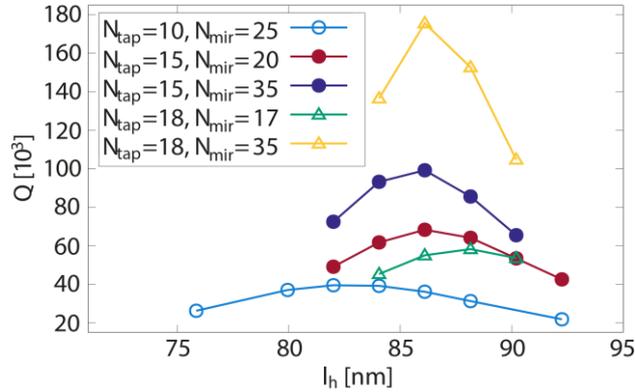

**Figure S1: Q-factor vs. defect length $l_h$ for the mode-matching PhC design.** Comparison of scattering-limited ($N_{mir} = 35$) and unsaturated ($N_{mir} = 35 - N_{tap}$) PhCs for different $N_{tap}$ values.

For the mode-matching design the resonance frequency as well as the Q-factor depends sensitively on the defect length $l_h$. We therefore optimize the defect length for different numbers of taper and mirror holes as shown in Figure S1.

**Fabrication of nanobeam PhC cavities**

We fabricate nanobeam photonic crystal cavities from commercial wafers with 200 nm stoichiometric $Si_3N_4$, on top of 2 µm thermal $SiO_2$ on Si layer-stacks. Nanophotonic devices, including the PhC cavities, are patterned in ma-N 2403 resist using electron beam lithography on a 100kV Raith EBPG 5150 system. After development in MF-319 and resist reflow the pattern is transferred into the SiN-layer by reactive ion etching using $CHF_3/O_2$ chemistry.

**Measurement**

Optical transmission measurements are performed by positioning chips with hundreds of nanophotonic devices under an array of 630 HP single mode fibers with 127 µm pitch. Translation stages with travel in XYZ-directions allow for accurate alignment of the fiber array with respect to a device under test (fiber-to-chip-interfaces are realized as optical grating couplers). The input fiber is connected via a polarization controller to a supercontinuum laser (NKT, SuperK COMPACT), whereas the output fiber connects to a spectrometer (RGB Qwave). Higher resolution spectra for accurate determination of quality factors are acquired with a Princeton Instruments 320PI spectrograph with a 2400 LP/mm grating, read out with a (liquid nitrogen cooled) Spec10 camera, yielding a resolution of <0.1 nm. For data collection, the position of the fiber-array and the polarization is optimized for maximal transmission at the resonance wavelength.

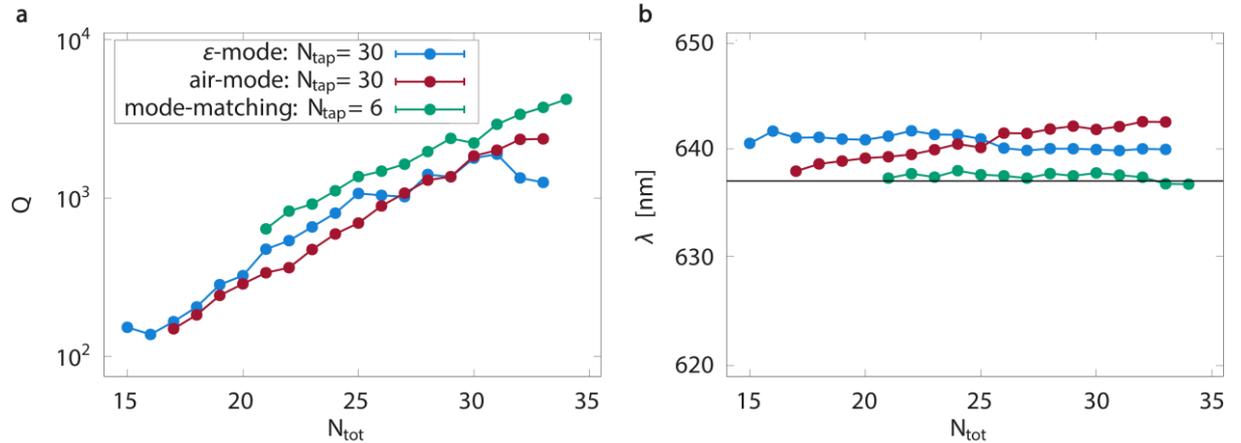

**Figure S2: Q-factor and resonance wavelength $\lambda$ vs. number of total holes.**
Variation of total number of holes $N_{tot} = N_{mir} + N_{tap}$ for the three PhC cavity design approaches.

For PhC cavity devices with optimized taper section, we determine if the cavity performance is limited by the overall reflectivity (number of holes) or by scattering and absorption (fabrication imperfections and material). With an increasing number of mirror holes, the reflectivity and the corresponding contribution to the Q-factor "$Q_{trans}$" increases, whereas the scattering contribution remains constant. Experimentally, we find an exponential increase of the Q-factor with the total number of holes for both deterministic and mode-matching designs in cavities with less than 25 mirror holes, as shown in Figure S2 a. The slope remains constant for the mode-matching design, but starts to decrease for the air-mode and dielectric mode geometries at ~ 30 and ~25 taper holes, respectively. This indicates that the scattering contribution to the Q-factor limits the performance of these cavities. We further observe that the resonance wavelengths stay within a 5 nm band of the target wavelength for all three designs as shown in Figure S2 b, which indicates constant mirror strength.